\documentclass[slac_one,letterpaper]{revtex4}
\usepackage{graphicx}
\usepackage{fancyhdr}
\def\met{\ensuremath{E_{\mathrm{T}}^{\mathrm{miss}}}}
\def\nb{nb$^{-1}$}
\begin{document}

\title{Early $W$ and $Z$ Measurements at ATLAS} 

%

\author{C. Mills, for the ATLAS collaboration}
\affiliation{Harvard University, Cambridge, MA 02138, USA}

\begin{abstract}
The first measurements of the $W$ and $Z$ cross sections in 
$pp$ collisions at $\sqrt{s} = 7$ TeV using the ATLAS detector at the 
Large Hadron Collider (LHC) have been completed.  Cross sections in the
electron and muon channels , as well as the combined cross section, are
presented for both the $W$ and the $Z$ boson.
The charge asymmetry of $W$ production as a function
of the psuedorapidity of the lepton has also been measured.  
\end{abstract}

\maketitle

\thispagestyle{fancy}


\section{INTRODUCTION} 
Observation of the $W$ and $Z$ bosons in $\sqrt{s} =$ 7 TeV proton-proton collisions
at the LHC is an important milestone in the ATLAS physics program.  These are
the first cross section measurements at 7 TeV that can be compared to a theoretical 
calculation of comparable precision.  Also, they are the most common 
source of isolated, high transverse momentum leptons, and are therefore
an important data sample for understanding final states based on leptons.

The $W$ and $Z$ bosons were first observed by the UA1 and UA2 experiments at CERN 
nearly thirty years ago, in proton-antiproton collisions at 
$\sqrt{s} = $540~GeV~\cite{Arnison:1983rp,Banner:1983jy,Arnison:1983mk,Bagnaia:1983zx}.
Since then, their properties have been characterized in detail by a succession
of collider experiments.  Now, confirmation of their known properties, such as their
masses and widths, and completion of collider-specific measurements,
such as their inclusive and differential cross sections, are a priority in the 
physics program of a new collider such as the LHC.

In the analyses described here, the $W$ decays to a charged lepton
and a neutrino, and the $Z$ decays to two charged leptons of opposite sign.  
The charged leptons may be either electrons or muons.  The isolated, energetic
leptons allow a very pure signal to be identified above the background.  
The observed data are compared to the signal and background expectations.
The signal acceptance times efficiency is calculated from Monte Carlo 
simulations, with corrections for differences between real and simulated
detector performance.  Background expectations are also derived primarily 
from simulation, with some use of data to make the QCD background expectations 
more robust.  This knowledge combined allows calculation of the production
cross sections for the $W$ and $Z$.  Using the same information, the lepton
charge asymmetry for $W$ production is also measured.

\section{THE ATLAS DETECTOR}
The ATLAS detector~\cite{DetectorPaper:2008} at the LHC 
consists of concentric cylindrical layers of inner tracking, calorimetry, and outer 
(or muon) tracking, with both the inner and outer tracking volumes contained in the 
fields of superconducting magnets to enable measurement of charged particle momenta.
In the ATLAS coordinate system, the $z$ axis points along the anti-clockwise beam 
direction, and the azimuthal and polar angles $\phi$ and $\eta$
are defined in the conventional way, with $\phi = 0$ (the $x$ axis) pointing from
the origin to the center of the LHC ring. The pseudorapidity is defined as 
\mbox{$\eta = -\ln \tan(\theta/2)$}. The transverse momentum $p_T$, the transverse 
energy $E_T$, and the transverse missing energy \met\ are defined in the $x-y$ plane.  

The Inner Detector (ID) provides precision tracking of charged particles inside of
$|\eta| \approx 2.5$.  It is located immediately around the interaction point, inside
a superconducting solenoid which produces a 2~T axial field.  The innermost layers use
silicon pixel and strip tracking technology, and the outer layers are a gaseous 
tracker which also provides transition radiation information.

The calorimeter system surrounds the ID and the solenoid.  It covers the pseudorapidity range 
$|\eta| < 4.9$ and has a minimum depth of 22 electromagnetic radiation lengths ($X_0$). 
The liquid argon (LAr) electromagnetic (EM) calorimeter uses a lead absorber in folded layers
designed to minimize gaps in coverage.  
The LAr calorimeter is segmented in depth to enable better particle shower reconstruction.
The innermost layer, or ``compartment'' is instrumented with 
strips that precisely measure the shower location in $\eta$.  The middle compartment 
is the deepest and contains most of the electromagnetic shower produced by a typical electron
or photon.  The outermost compartment has the coarsest granularity and is used to quantify
how much of the particle shower is leaking back into the hadronic calorimeter.

The outermost tracking layers, interleaved with superconducting air-core toroid magnets, 
form a muon spectrometer.  Precision tracking in
the bending plane ($R-\eta$) for both the barrel and the endcaps is done by 
drift tubes.  In the innermost forward plane, from $2.0 < |\eta| < 2.7$, 
precision spacepoints are provided by cathode strip chambers.   
Resistive plate chambers and thin gap chambers are used in the barrel and endcap
muon triggers, respectively.  In addition to fast readout for 
triggering, these detectors provide $\phi$ hit information for offline muon reconstruction.

\section{\label{SampleSection}CANDIDATE DATA SAMPLES AND MODELING OF SIGNAL AND BACKGROUND}
The $W$ cross section and asymmetry measurements shown here use approximately
17 \nb\ of integrated luminosity.  Electron events are selected online through a 
hardware-based trigger that selects events with an electromagnetic cluster with $E_T > 5$ 
GeV and $|\eta| < 2.5$.  Muon events are selected online by comparing the pattern
of hits in the muon spectrometer to a lookup table based on muon tracks.
The $W$ analysis uses a muon trigger with ``open roads'', that is, no explicit $p_T$ 
threshold.  

The $Z$ cross section measurement uses 219--229 \nb of integrated luminosity.  
Because the additional data used was acquired after the data for the $W$ analysis,
higher instantaneous luminosity conditions made it necessary to raise the
trigger thresholds, to 10 GeV for the electron channel and 6 GeV for the muon channel.  

The signal and backgrounds are modeled using {\sc Pythia}~\cite{pythia} Monte 
Carlo using MRSTLO*~\cite{mrst} parton distribution functions (PDF)
integrated with a {\sc Geant4}~\cite{geant4} simulation of the ATLAS detector.

\section{\label{SelectionSection}EVENT SELECTION}
Candidate $W$ and $Z$ events are identified based on selection of events with
at least one lepton.  Additional selection uses the second lepton (neutrino 
or charged) and the mass of the boson to enhance the signal to background 
ratio.

\subsection{Preselection}
\begin{figure*}[t]
\centering
\includegraphics[width=0.4\textwidth]{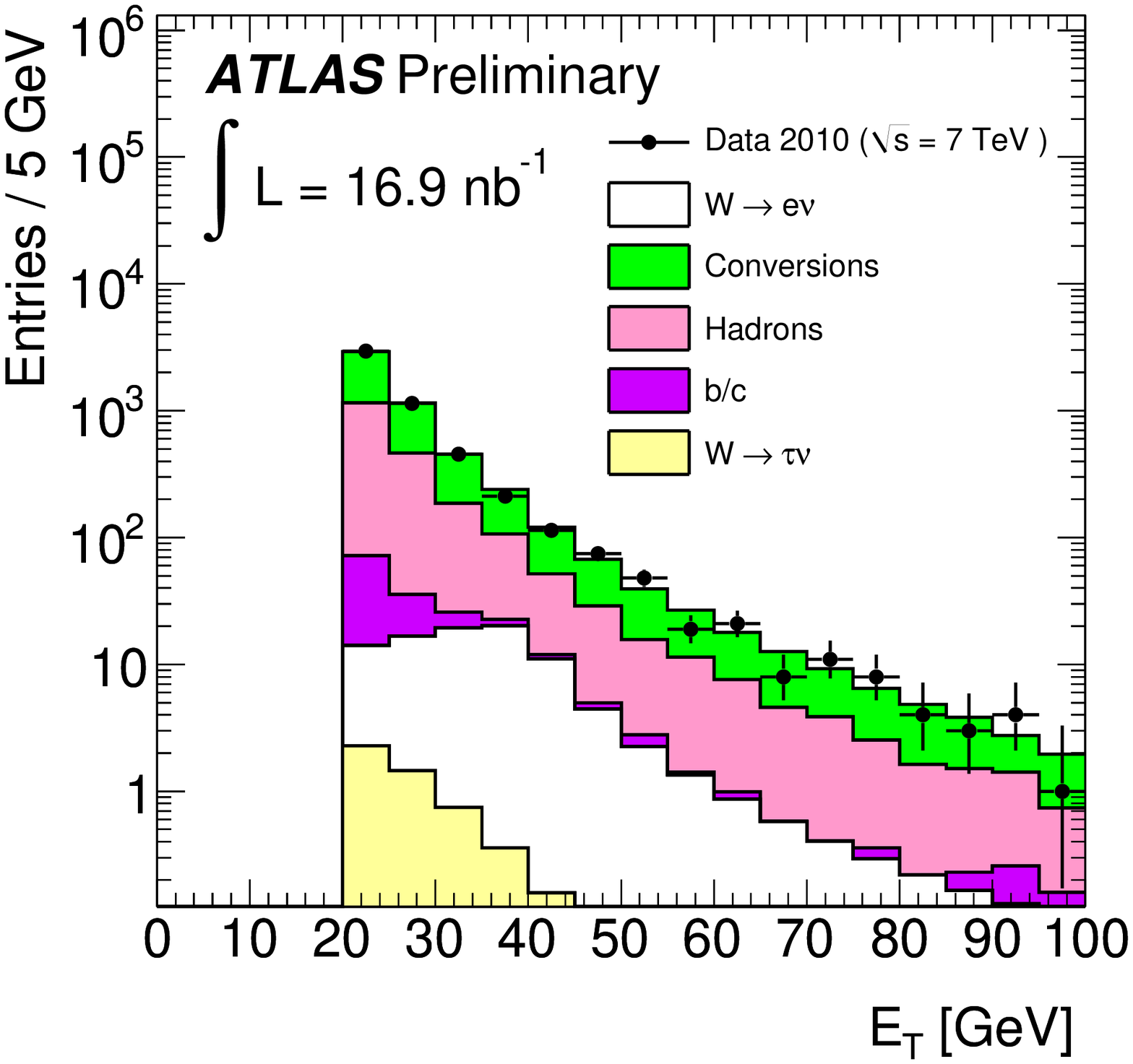}
\includegraphics[width=0.4\textwidth]{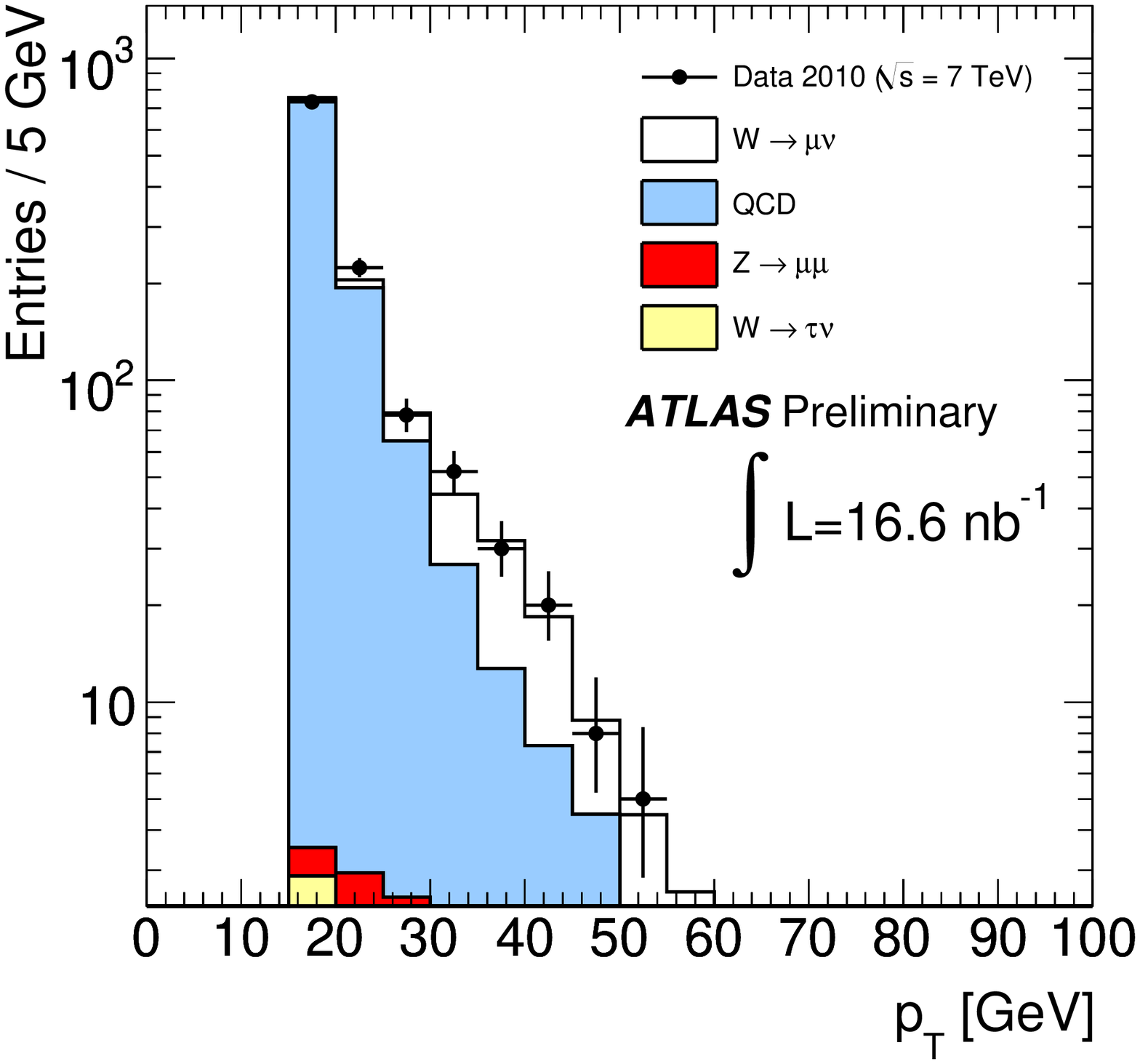}
\caption{\label{Preselection}Transverse energy of electrons passing preselection
  (left) and transverse momentum of muons passing preselection (right).} 
\end{figure*}

Electrons at ATLAS are reconstructed as a calorimeter cluster matched to
an Inner Detector track, with transverse energy $E_T > 20$~GeV and 
$|\eta| < 2.47$, excluding the range from $1.37 < |\eta| < 1.52$.  
Electron selection at ATLAS has three tiers of increasing stringency:
``loose'', ``medium'', and ``tight''.  The ``loose'' selection uses information
from the middle compartment of the LAr calorimeter to require that the shower profile
is consistent with an electron.  The ``medium'' selection
adds the requirement of a high-quality matched ID track.  It also uses the fine-granularity
information from the innermost calorimeter layer for additional checks of the 
shower shape.  ``Tight'' electrons pass additional track quality and kinematic requirements.
More detail on electron identification is available in Ref.~\cite{eg900}.  
For the event preselection, all electrons passing the ``loose'' selection are considered.
The $E_T$ of electron candidates satisfying the preselection criteria 
in events passing the 5 GeV trigger are shown in Figure~\ref{Preselection}.

Muons are reconstructed by combining an inner detector (ID) and muon spectrometer (MS)
track, and must have $p_T > 15$ GeV and $|\eta| \leq 2.4$ are to pass the preselection.  
To reduce the background from cosmic rays, the $z$ position where a 
candidate muon passes the beamline must be within 1~cm of the $z$ of a 
reconstructed primary vertex.  Additionally, to reduce the background from
decays-in-flight, the MS track of the muon is required to have $p_T > 10$~GeV
and the absolute difference between the $p_T$ measured in the MS and the ID is 
required to be less than 15~GeV.  The $p_T$ 
of all preselected muons in events passing the ``open roads'' trigger
are shown in Fig.~\ref{Preselection}.

\subsection{$W$ Event Selection}
First, extra criteria are added to the lepton selection.
Electron candidates are required to pass the ``tight'' requirements, which
have the best QCD rejection.  In the muon channel, the $p_T$ threshold is 
raised to 20~GeV from 15~GeV and a track isolation requirement is added 
to reduce backgrounds, particularly from QCD.  

\begin{figure*}[t]
\centering
\includegraphics[width=0.4\textwidth]{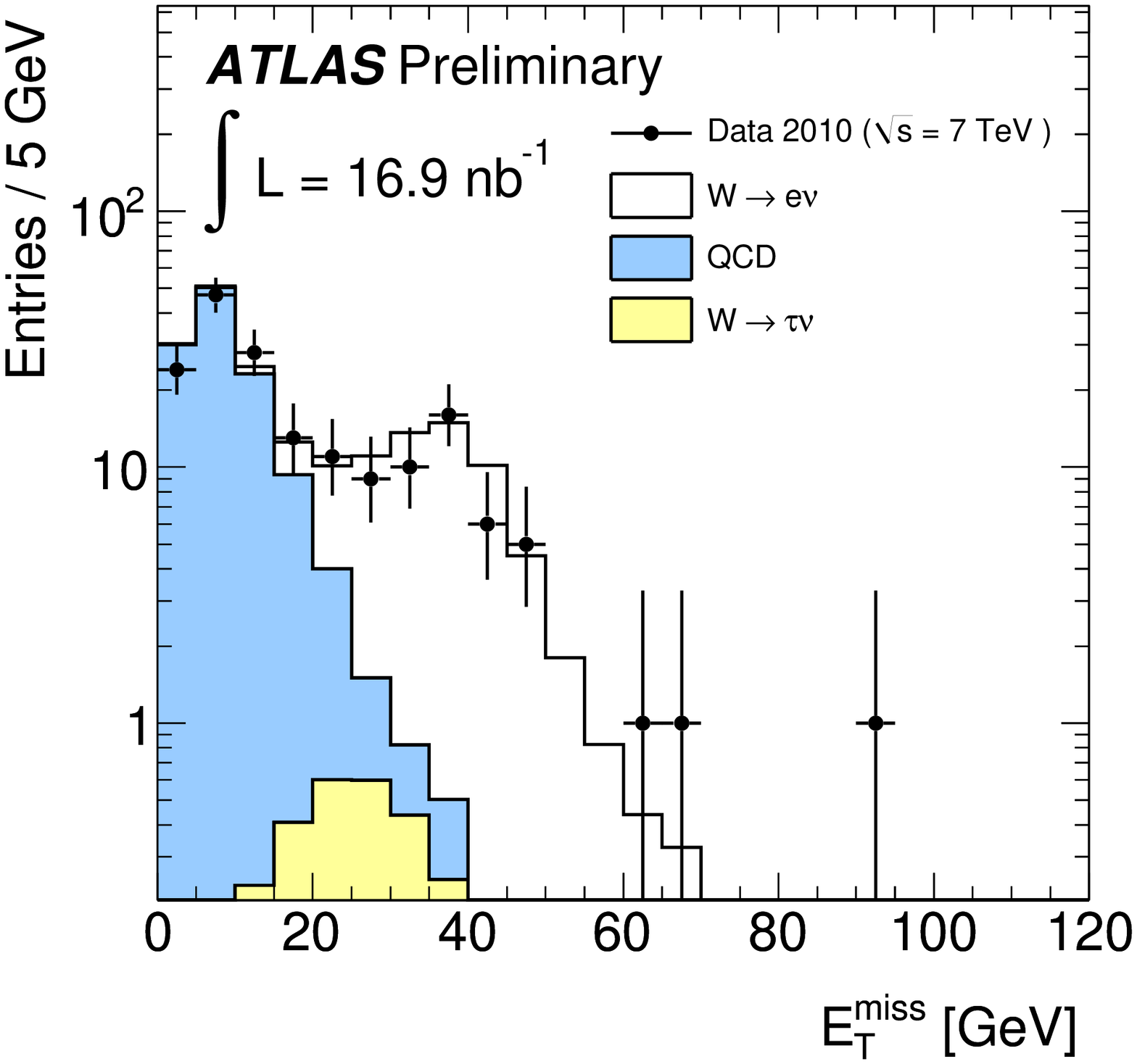}
\includegraphics[width=0.4\textwidth]{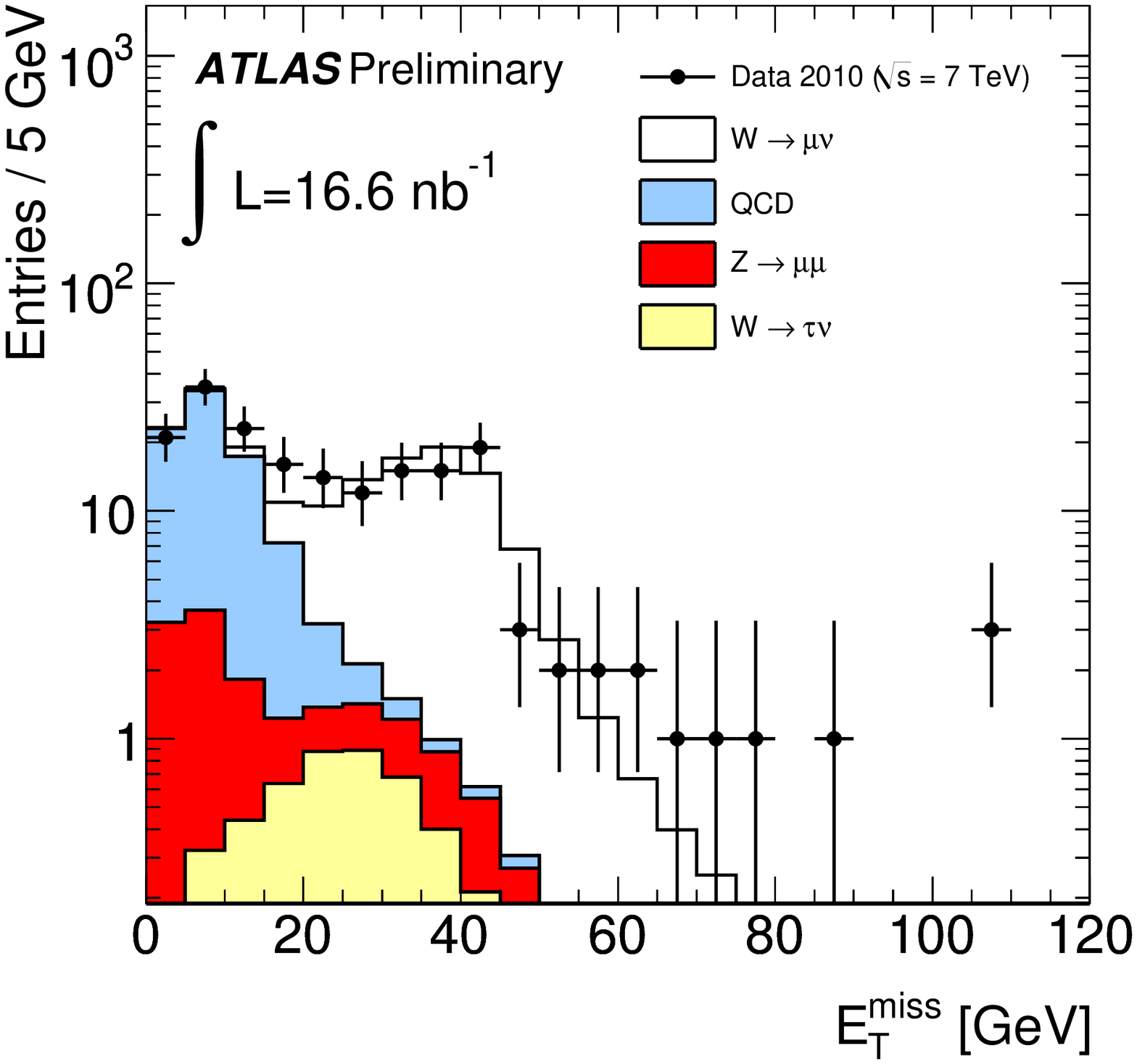}
\caption{\label{EtMiss} Missing transverse energy (\met) in events
passing all lepton selection for $W$ candidates in the electron (left)
and muon (right) channels.} 
\end{figure*}

\begin{figure*}[t]
\centering
\includegraphics[width=0.4\textwidth]{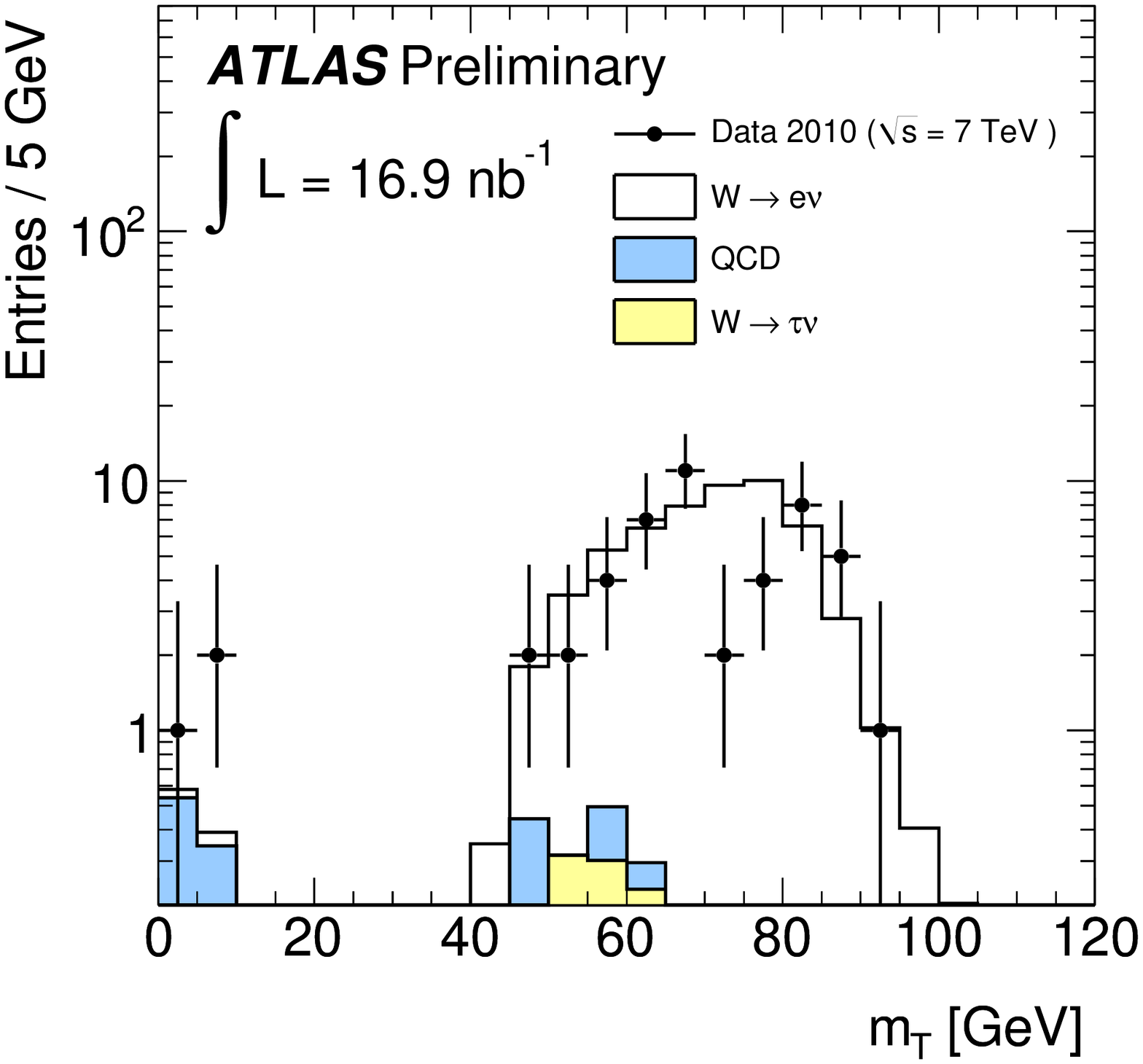}
\includegraphics[width=0.4\textwidth]{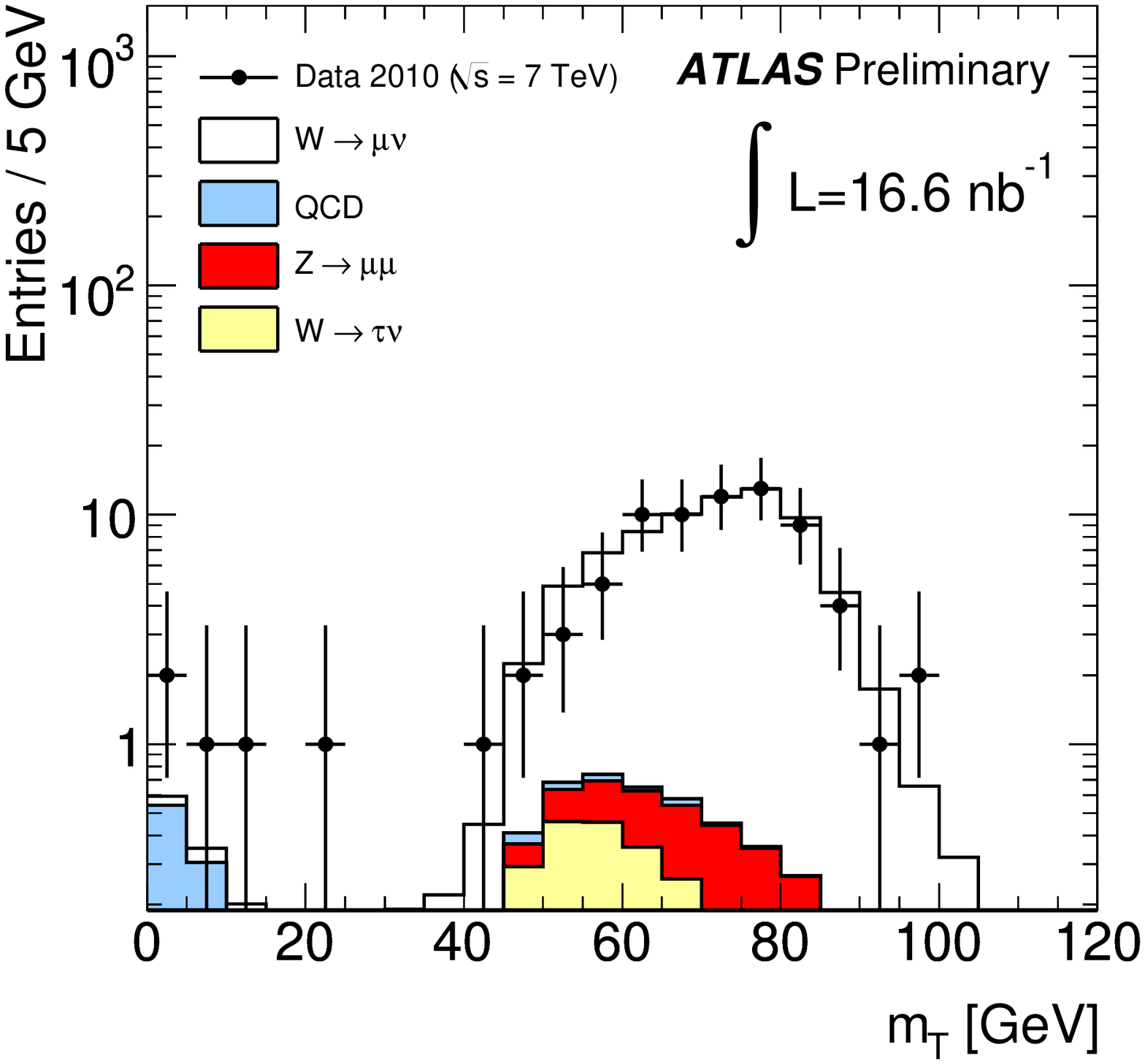}
\caption{\label{TransverseMass} Transverse mass of $W$ candidates in
the electron (left) and muon (right) channels after all event selection
except for the transverse mass requirement.} 
\end{figure*}

The missing transverse energy in $W\to\ell\nu$ events measures the transverse
momentum of the neutrino produced by the $W$ decay.  The \met\ in electron 
and muon events after the final lepton selection described in the previous 
paragraph is shown in Figure~\ref{EtMiss}.  The $W$ event selection requires
\met\ $>$ 25 GeV.
The transverse mass, defined as 
\[
M_T = \sqrt{2(p_T^\mu)(\met)(1 - \cos(\phi^\mu - \phi^{\met}))} \ ,
\]
is a distinctive property of $W$ events.  The transverse mass distribution
of candidates in the electron and muon channels after the \met\ requirement
is shown in Fig.~\ref{TransverseMass}.  A requirement of \mbox{$M_T > 40$ GeV}
completes the $W$ event selection for both channels.

\subsection{$Z$ Event Selection}
\begin{figure*}[t]
\centering
\includegraphics[width=0.4\textwidth]{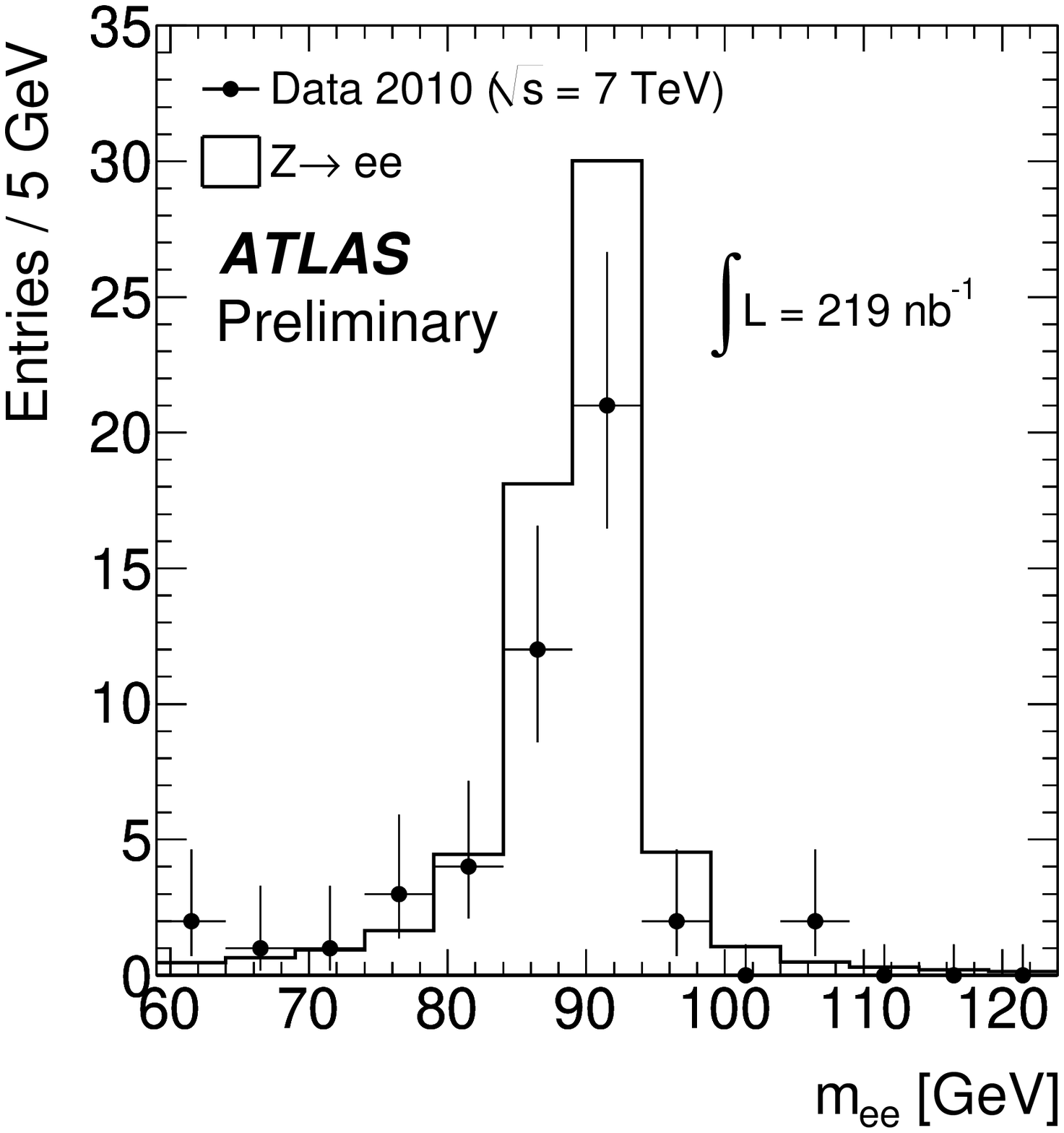}
\includegraphics[width=0.4\textwidth]{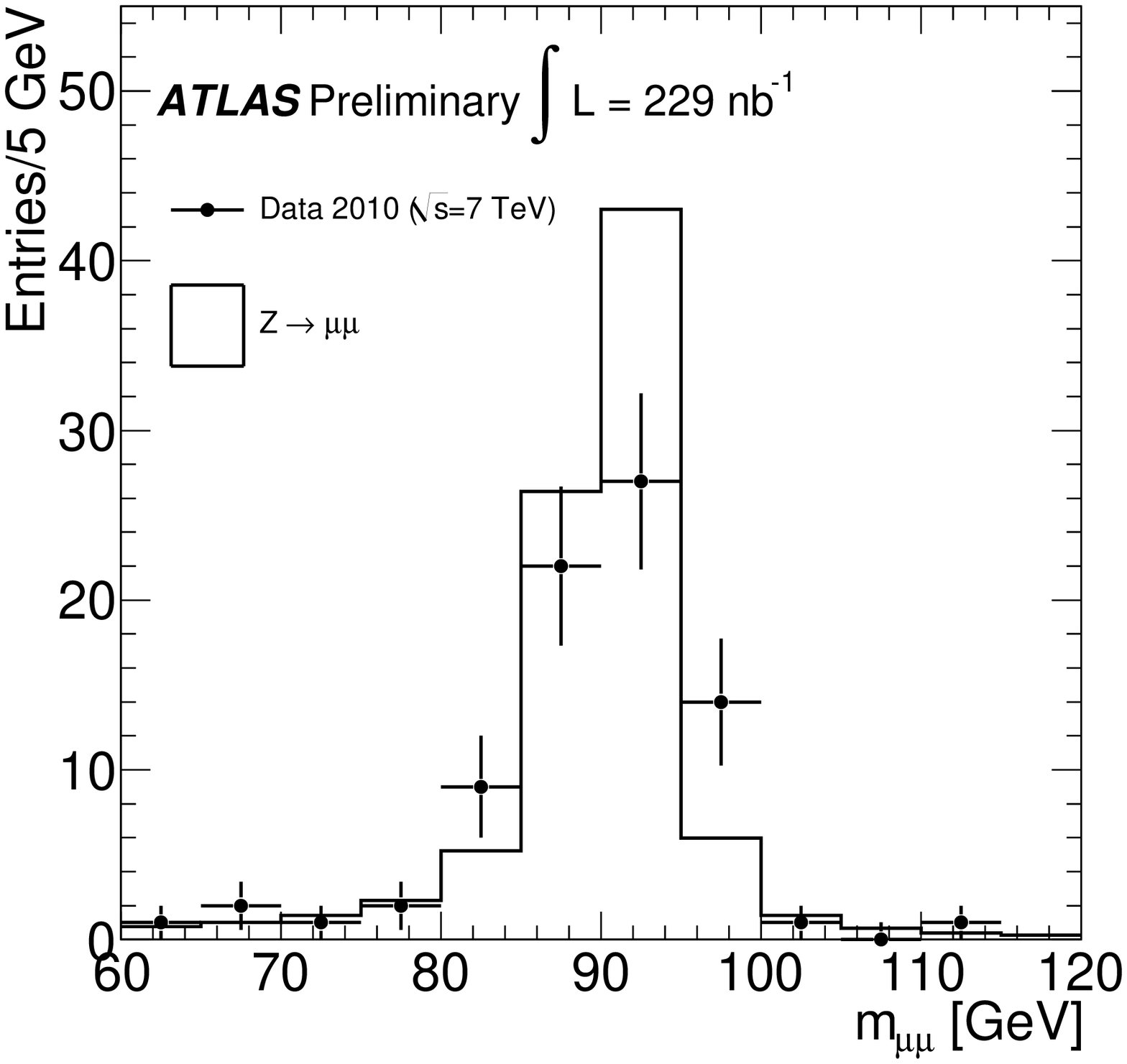}
\caption{\label{InvariantMass} Transverse mass of $W$ candidates in
the electron (left) and muon (right) channels after all event selection
except for the transverse mass requirement.} 
\end{figure*}

For both the electron and muon channels, the $Z$ event selection requires two
same-flavor oppositely charged leptons with an invariant mass reconstructed in the
range \mbox{$66 < M_{\ell\ell} < 116$ GeV}.  In the electron channel, the electron
selection is as for the $W$ analysis, except that the electrons are only required 
to pass the ``medium'' selection, increasing the acceptance.
Muons are selected identically to the $W$ analysis.
The invariant mass distribution of $Z$ candidates immediately before the $M_{\ell\ell}$
requirement is shown in Fig.~\ref{InvariantMass}.

\section{\label{WAcceptanceSection}$W$ ACCEPTANCE AND ASSOCIATED SYSTEMATIC UNCERTAINTIES}
The $W\to\ell\nu$ acceptance times efficiency can
be factorized into a geometrical and kinematic acceptance $A_W$ and a correction
factor $C_W$ for reconstruction efficiency and resolution effects.  
In the analysis presented here, both are calculated from {\sc Pythia}, although
$C_W$ is corrected for observed discrepancies between true and simulated detector
performance where possible.
Table~\ref{WAcceptance} shows $A_W$, $C_W$, and their product, the total 
acceptance times efficiency, for both the electron and muon channels.
\begin{table}[t]
\begin{center}
\caption{Components of the $W\to\ell\nu$ acceptance times efficiency, with associated
systematic uncertainties.}
\begin{tabular}{|l|c|c|c|}
\hline 
\textbf{channel}	& $A_W$ 	& $C_W$		& \textbf{acceptance x efficiency} \\
\hline 
electrons	& 0.462 $\pm$ 0.014	& 0.656 $\pm$ 0.053	& 0.303 $\pm$ 0.026	\\
\hline 
muons		& 0.480 $\pm$ 0.014	& 0.814 $\pm$ 0.056	& 0.391 $\pm$ 0.029	\\
\hline 
\end{tabular}
\label{WAcceptance}
\end{center}
\end{table}

The acceptance $A_W$ is defined as the fraction of all generated events
passing the following criteria at truth level: $p_T^{\ell}>20$~GeV, 
$p_T^{\nu}>25$~GeV, $M_T>40$~GeV, as well as  
$|\eta_e| < 1.37$ or $1.52 < |\eta_e| < 2.47$ or $|\eta_\mu| < 2.4$.
The systematic uncertainty on the acceptance $A_W$, 3\%, is 
common between the electron and muon channels.  The dominant contribution is 
from differences between the {\sc Pythia} acceptance calculation and the
acceptance as calculated using the MC@NLO~\cite{mcatnloDYpatch} generator.
The uncertainty also includes the differences observed in the acceptance
when the PDF set used is varied.

The correction factor $C_W$ is the ratio of the number of events passing
all of the $W\to\ell\nu$ event selection after full reconstruction to the 
number of events passing the truth-level selection.
The systematic uncertainty on $C_W$ includes the uncertainties
on the trigger and reconstruction efficiencies, as well as the uncertainty
on the lepton and \met\ energy scale and resolutions.  The total is 8\% for the 
electron channel and 7\% for the muon channel.  

\section{$Z$ ACCEPTANCE AND ASSOCIATED SYSTEMATIC UNCERTAINTIES}

The $Z\to\ell\ell$ acceptance times efficiency is similarly factorized into a
geometrical and kinematic component $A_Z$ and a correction for detection efficiency
$C_Z$.  These components, and their product, the total acceptance times efficiency,
are shown together with their associated uncertainties in Table~\ref{ZAcceptance}.
\begin{table}[t]
\begin{center}
\caption{Components of the $Z\to\ell\ell$ acceptance times efficiency, with associated
systematic uncertainties.}
\begin{tabular}{|l|c|c|c|}
\hline 
\textbf{channel}	& $A_Z$ 	& $C_Z$		& \textbf{acceptance x efficiency} \\
\hline 
electrons	& 0.446 $\pm$ 0.013	& 0.645 $\pm$ 0.090	& 0.288 $\pm$ 0.091	\\
\hline 
muons		& 0.486 $\pm$ 0.014	& 0.797 $\pm$ 0.055	& 0.387 $\pm$ 0.057	\\
\hline 
\end{tabular}
\label{ZAcceptance}
\end{center}
\end{table}

The geometrical and kinematic acceptance $A_Z$ is the fraction of all generated 
events with $66 < m_{\ell\ell} < 116$~GeV which have two leptons with $p_T > 20$~GeV
and $|\eta_e| < 1.37$ or $1.52 < |\eta_e| < 2.47$ or $|\eta_\mu| < 2.4$.
The 3\% systematic uncertainty on $A_Z$ consists of approximately equal contributions
from PDF uncertainties and LO-NLO differences.
The efficiency correction factor $C_Z$ is the ratio of the number of events passing
all of the $Z$ event selection to the number passing the truth-level selection 
described above.  Because there are two charged leptons in the final state,
the systematic uncertainty on $C_Z$ is dominated by uncertainties on
lepton trigger and reconstruction efficiencies in both channels.  In the electron
channel, the total uncertainty is 14\%, and for muons, it is 7\%.

\section{\label{BackgroundSection}BACKGROUND CALCULATIONS}

\subsection{Backgrounds to $W\to\ell\nu$}
For the $W\to\ell\nu$ analyses, the $Z/\gamma^* \to \ell\ell$ and $W\to\tau\nu$
backgrounds are estimated using acceptances calculated from {\sc Pythia} Monte
Carlo simulations.  The systematic uncertainties on these backgrounds are 
calculated similarly to the systematic uncertainties on the $W$ acceptance times 
efficiency.  

In the electron channel, the QCD background is calculated from a fit of the 
isolation distribution of candidate events selected using
``loose'' instead of ``tight'' electrons with $W\to e\nu$ and QCD templates 
from simulation.  The number of QCD events 
resulting from the fit is scaled by a rejection factor, measured in simulation,
for the tight selection relative to the loose selection.  
In the muon channel, the number of isolated, low-\met\ QCD events is extrapolated 
to the number of isolated, high-\met\ QCD events by multiplying it by the ratio 
of the number of non-isolated, high-\met\ events to the number of non-isolated, 
low-\met\ events.  
The uncertainties on both QCD background estimates contain a significant statistical
contribution.  The systematic uncertainties on the QCD backgrounds
are based on variations of the methods used for the electron channel and the 
results of a closure test in simulation for the muon channel.  

\subsection{Backgrounds to $Z\to\ell\ell$}
Backgrounds to $Z\to\ell\ell$ are very small (1~\% or less) relative to the signal.  For both 
the electron and muon channels, contributions from $t\bar{t}$, $Z\to\tau\tau$,
and $W\to\ell\nu$ are calculated from simulation, and their associated systematics
are calculated as for the signal acceptance times efficiency.  For the muon 
channel, the QCD background is calculated in the same way.  For 
the electron channel, the QCD background is based on the number of events passing 
the $Z\to ee$ selection in simulation, but with two loose electrons instead of two medium 
electrons.  Then, a measurement in data of the 
medium-to-loose rejection for electrons from QCD is used to scale this number to 
the number of expected events where both electrons pass the medium identification 
requirements.

The predictions of 0.49~$\pm$~0.09 and 0.17~$\pm$~0.01 background events for the
electron and muon channels compares favorably with the fact that one same-sign
two-electron event and no same-sign muon events are observed passing all of the 
other $Z$ candidate selection.

\section{CROSS SECTION RESULTS}
After all of the $W$ event selection, 46 candidate events are observed in the electron
channel and 72 are observed in the muon channel.  Combining with the integrated 
luminosity $\int L\;\mathrm{d}t$ of 
the sample, the acceptance times efficiency $A_W \times C_W$ 
and the background estimates $N_{\textrm{background}}$,
the cross section can be calculated using the formula
\[
\sigma = \frac{N_{\textrm{cand}} - N_{\textrm{background}}}{A_W \times C_W \times \int L\;\mathrm{d}t}\ .
\]
The $W$ cross section input and results for the electron and muon channels are presented,
along with the combined result, in Table~\ref{WCrossSection}.  
The uncertainties on the cross sections include the 11\% uncertainty on the luminosity~\cite{lumiconf}.
The results are all in agreement with the NNLO standard model prediction of 10.46~$\pm$~0.02~nb,  
as calculated using the FEWZ program~\cite{fewz1, fewz2} with the MSTW2008 parton distribution
functions~\cite{Martin:2009iq}.
\begin{table}[t]
\begin{center}
\caption{Results for the $W$ cross section times branching ratio to leptons, with inputs.}
\begin{tabular}{|l|c|c|c|c|r|}
\hline 
\textbf{channel}	& \textbf{int.~lum. (nb$^{-1}$)}	& $N_\textrm{cand}$	& $N_{\textrm{background}}$	& \textbf{acceptance x efficiency} & $\sigma \times$ \textbf{BR (nb)}\\
\hline 
electron	& 16.9	& 46	& 2.6 $\pm$ 0.5	& 0.303 $\pm$ 0.026	&
8.5 $\pm$ 1.3 (stat) $\pm$ 0.7 (sys) $\pm$ 0.9 (lum) \\
\hline 
muon		& 16.6	& 72 	& 5.3 $\pm$ 0.7	& 0.391 $\pm$ 0.029	&
10.3 $\pm$ 1.3 (stat) $\pm$ 0.8 (sys) $\pm$ 1.1 (lum) \\
\hline 
combined	& -	& 118	& -	& - & 9.3 $\pm$ 0.9 (stat) $\pm$ 0.6 (sys) $\pm$ 1.0 (lum) \\
\hline 
\end{tabular}
\label{WCrossSection}
\end{center}
\end{table}

The $Z$ cross section inputs and results are shown in Table~\ref{ZCrossSection}.  There are
46 candidate events passing all event selection in the electron channel and 79 in the muon
channel.  The cross section times branching ratio is calculated using an equivalent formula to 
the one used for the $W$ result.  The predicted cross section times branching ratio for 
66~GeV~$< M_{\ell\ell} <$~116~GeV is 0.964~$\pm$~0.039~nb, also calculated in the same way as 
for the $W$.
\begin{table}[t]
\begin{center}
\caption{Results for the $Z$ cross section times branching ratio to leptons, with inputs.}
\begin{tabular}{|l|c|c|c|c|r|}
\hline 
\textbf{channel}	& \textbf{int.~lum. (nb$^{-1}$)}	& $N_\textrm{cand}$	& $N_{\textrm{background}}$	& \textbf{acceptance x efficiency} & $\sigma \times$ \textbf{BR (nb)}\\
\hline 
electron	& 219	& 46	& 0.49 $\pm$ 0.09	& 0.288 $\pm$ 0.091	&
0.72 $\pm$ 0.11 (stat) $\pm$ 0.10 (sys) $\pm$ 0.08 (lum) \\
\hline 
muon		& 229	& 79 	& 0.17 $\pm$ 0.01	& 0.387 $\pm$ 0.057	&
0.89 $\pm$ 0.10 (stat) $\pm$ 0.07 (sys) $\pm$ 0.10 (lum) \\
\hline 
combined	& -	& 125	& -	& - & 0.83 $\pm$ 0.07 (stat) $\pm$ 0.06 (sys) $\pm$ 0.09 (lum) \\
\hline 
\end{tabular}
\label{ZCrossSection}
\end{center}
\end{table}

\section{$W$ CHARGE ASYMMETRY}
\begin{figure*}[t]
\centering
\includegraphics[width=0.45\textwidth]{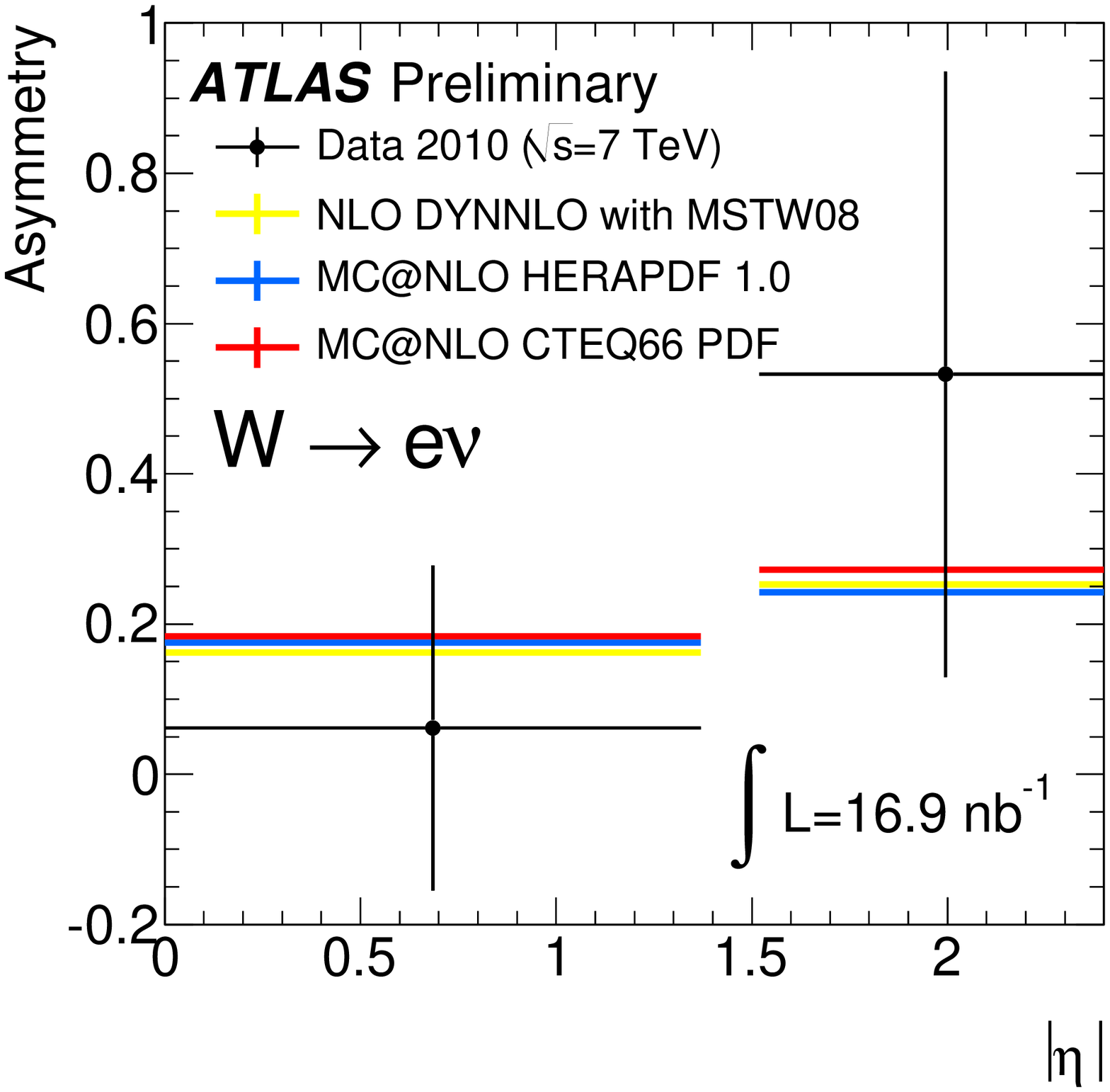}
\includegraphics[width=0.45\textwidth]{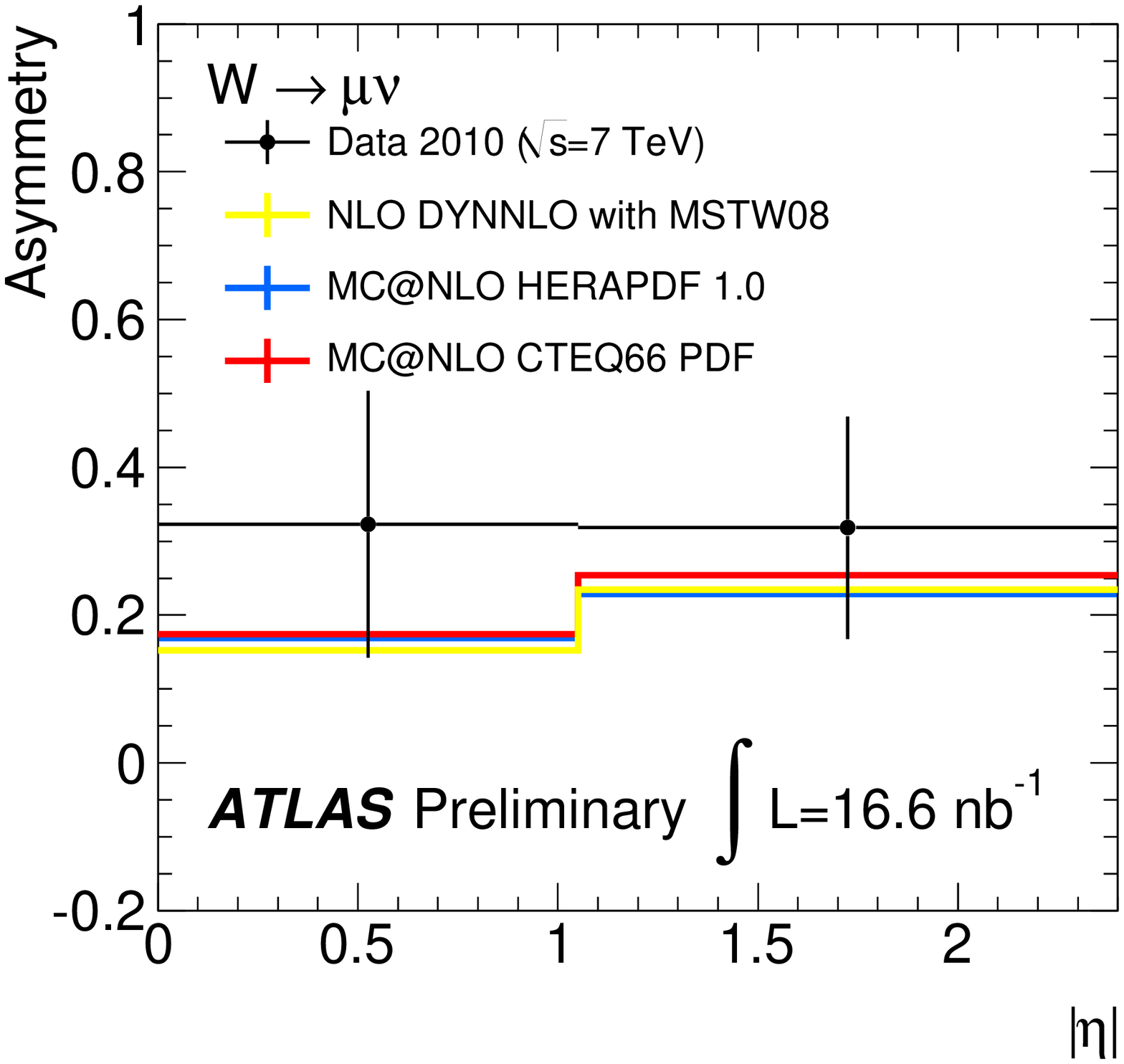}
\caption{\label{WAsymmetryEta} Lepton charge asymmetries in $W$ production as
a function of lepton pseudorapidity in the electron (left) and muon (right) channels,
compared to three theoretical predictions (see text).}
\end{figure*}

One distinctive feature of $W$ production at the LHC is that, because it is a proton-proton
collider, more positive than negative $W$s are produced.  Also, the asymmetry depends on the 
momentum fraction carried by the colliding quarks and antiquarks, so the ratio of the number 
of $W^+$ to the number of $W^-$ produced depends on the lepton pseudorapidity.   Using the 
$W$ sample selected for the cross section measurement, it is possible to measure the 
charge asymmetry, defined as
\[
A = \frac{\sigma^{\ell^+}-\sigma^{\ell^-}}{\sigma^{\ell^+}+\sigma^{\ell^-}} \ ,
\]
where $\sigma^{\ell^{+(-)}}$ is the cross section measured with positive (negative) leptons.
The fiducial cross sections, defined with $A_W$ set to 1.0, are used to measure the 
asymmetry, reducing the dependence of the result on theoretical assumptions.
The results for the electron and muon channels in two bins of $|\eta|$ are shown in 
Figure~\ref{WAsymmetryEta}.  The integral results are 0.21~$\pm$~0.18~(stat.)~$\pm$0.01~(sys.)
in the electron channel and 0.33~$\pm$~0.12~(stat.)~$\pm$~0.01~(sys.) in the muon channel.  These may be compared
to the NLO theoretical prediction, 0.20, which is calculated using the DYNNLO~\cite{Catani:2007vq}
program with the MSTW08 PDF set~\cite{Martin:2009iq}.  Figure~\ref{WAsymmetryEta} includes that prediction
as well as ones from MC@NLO~\cite{mcatnloDYpatch} interfaced with the CTEQ6.6~\cite{Nadolsky:2008zw} 
and HERAPDF 1.0~\cite{HERAPDF} PDF sets.

\section{SUMMARY}
The distinctive signatures of $W$ and $Z$ production have been observed at the
ATLAS detector at the LHC.  First cross section measurements have been performed, and
the results are in overall good agreement with the Standard Model predictions.
Furthermore, the expected charge asymmetry in $W$ production, a unique feature
of $W$ production in proton-proton collisions, has been observed and measured.
These first measurements will pave the way for a full program of measurements of
the $W$ and $Z$ at the LHC, which will deepen our understanding of the Standard 
Model as well as establishing key tools, such as lepton identification, for use
in the detection of rare and new processes.



\end{document}